\newcommand{\version}{May 4, 2010}
\documentclass[11pt,a4paper,english]{article}

\usepackage[bindingoffset=0.35cm,textheight=22.5cm,hdivide={2.7cm,*,2.7cm}, vdivide={*,22cm,*}]{geometry}
\usepackage[dvips]{graphicx}
\usepackage[dvips,bookmarksnumbered=true,breaklinks=true]{hyperref}

\usepackage{amsmath,amsfonts,amssymb,babel}
\usepackage{dsfont}

\usepackage[numbers,square,comma,sort&compress]{natbib}

\newcommand{\startappendix}{\appendix}

\newcommand{\Appendix}[1]{
\vspace{2.7em}
\pagebreak[3]
\refstepcounter{section}
\begin{flushleft}
{\Large\bf Appendix \thesection:\hspace*{1ex} #1}
\addcontentsline{toc}{section}{Appendix \thesection}
\end{flushleft}}

\newcommand{\qed}{\nobreak \ifvmode \relax \else
      \ifdim\lastskip<1.5em \hskip-\lastskip
      \hskip1.5em plus0em minus0.5em \fi \nobreak
      \vrule height0.75em width0.5em depth0.25em\fi}

\newcommand{\be}{\begin{equation}}
\newcommand{\ee}{\end{equation}}
\newcommand{\eq}[1]{(\ref{#1})}

\def\nn{\nonumber}
\def\bea{\begin{eqnarray}}
\def\eea{\end{eqnarray}}

\def\beqa{\begin{eqnarray}} 
\def\eeqa{\end{eqnarray}} 
\def\beq{\begin{equation}} 
\def\eeq{\end{equation}}

\def\Tr{{\rm Tr}}

\def\a{\alpha}          
\def\b{\beta}           

\def\e{\epsilon}                
        
\def\g{\gamma}

\def\m{\mu}     \def\n{\nu}

\def\r{\rho}
\def\s{\sigma}  

\def\th{\theta}  \def\vth{\vartheta}

 \def\cH{{\cal H}} 
\def\cJ{{\cal J}}  \def\cL{{\cal L}}
\def\cM{{\cal M}}

\newcommand{\R}{\mathds{R}}

\def\bit{\begin{itemize}}
\def\eit{\end{itemize}}

\def\({\left(}
\def\){\right)}
\def\diag{\mbox{diag}}

\def\pa{\partial} \def\del{\partial}

\def\bcomment#1{}

\newcommand{\nc}{non-com\-mu\-ta\-tive}
\newcommand{\etal}{{\it et al.}}
\newcommand{\eqnref}[1]{Eqn.~(\ref{#1})}		
\newcommand{\secref}[1]{Section~\ref{#1}}		
\newcommand{\appref}[1]{Appendix~\ref{#1}}		
\newcommand{\inv}[1]{\frac{1}{#1}}				
\newcommand{\tinv}[1]{\tfrac{1}{#1}}
\newcommand{\pb}[2]{\{#1,#2\}}						
\newcommand{\co}[2]{[#1,#2]}						
\newcommand{\aco}[2]{[#1,#2]_+}						
\newcommand{\starco}[2]{\left[ #1\stackrel{\star}{,}#2\right] }		
\newcommand{\staraco}[2]{\left[ #1\stackrel{\star}{,}#2\right]_+ }	
\newcommand{\intg}{\int\!\sqrt{g}\,}					
\newcommand{\lpa}{\overleftarrow{\pa}}					
\newcommand{\rpa}{\overrightarrow{\pa}}
\newcommand{\mO}[1]{\mathcal{O}\left(#1\right)}				

\newcommand{\vph}{\varphi}

\newcommand{\w}{\omega}

\renewcommand{\Xi}{\Xi}

\newcommand{\W}{\Omega}

\newcommand{\syf}{\varTheta}	
\newcommand{\const}{\text{const.}}	
\newcommand{\cst}{c_1}		
\newcommand{\sth}{\epsilon}		
\newcommand{\bs}{{\bar{\sigma}}}	
\newcommand{\Tt}{\tilde{t}}

\title{\begin{flushright}
       \small{UWThPh-2010-07}
       \end{flushright}
\vspace{3em}
Schwarzschild Geometry Emerging from Matrix Models
}
\author{Daniel N. Blaschke\footnote{daniel.blaschke@univie.ac.at}~, Harold Steinacker\footnote{harold.steinacker@univie.ac.at}}

\date{\version}
\graphicspath{{./figures/}}
\begin{document}
\maketitle

\begin{center}
\renewcommand{\thefootnote}{\fnsymbol{footnote}}
\textit{Faculty of Physics, University of Vienna\\
Boltzmanngasse 5, A-1090 Vienna (Austria)}
\vspace{0.5cm}
\end{center}%
\begin{abstract}
We demonstrate how various geometries can emerge from Yang-Mills type matrix models with branes, and consider the examples of Schwarzschild and Reissner-Nordstr{\"o}m geometry. 
We provide an explicit embedding of these branes in $\R^{2,5}$ and $\R^{4,6}$, as well as an appropriate 
Poisson resp. symplectic structure which determines the non-commutativity of space-time. The embedding 
is asymptotically flat with asymptotically constant $\theta^{\mu\nu}$ for large $r$, and therefore suitable for a generalization to many-body configurations.
This is an illustration of our previous work~\cite{Blaschke:2010rg}, where we have shown how the Einstein-Hilbert action can be realized within such matrix models. 
\end{abstract}

\newpage
\tableofcontents

\section{Background and introduction}\label{sec:background}

It has been argued in numerous publications that combining the basic aspects of Quantum Mechanics and General Relativity 
strongly suggests a quantum structure of space-time itself near the Planck 
scale --- see for example Ref.~\cite{Doplicher:1994tu}. 
One approach to realize this idea is to replace classical space-time by a quantized, or {\nc} (NC), space-time. Coordinate functions 
$x^\m$ are promoted to Hermitian operators $X^\m$ acting on a Hilbert space $\cH$, which satisfy certain non-trivial 
commutation relations
\begin{align}
\co{X^\m}{X^\n}&= i\th^{\m\n}\,.
\end{align}
In the simplest case one may consider a Heisenberg algebra, corresponding to constant $\th^{\m\n}$
which commutes with the $X^\mu$. 
This has been studied extensively 
in the past (cf.~\cite{Douglas:2001ba,Szabo:2001kg,Rivasseau:2007a} for a review of such ``{\nc}'' field theories). 
However, in the context of gravity it seems essential that this commutator $\th^{\m\n}$ becomes dynamical. Indeed, semi-classically 
it determines a Poisson structure on space-time, as we will discuss below.

It has been shown previously~\cite{Steinacker:2007dq,Grosse:2008xr,Klammer:2008df} that  
 matrix models of Yang-Mills type naturally realize this idea, and
incorporate at least some version of (quantized) gravity; 
see Ref.~\cite{Steinacker:2010rh} for a review. Hence we start our discussion 
with the matrix model action
\begin{align}
S_{YM}&=-\Tr\co{X^a}{X^b}\co{X^c}{X^d}\eta_{ac}\eta_{bd}\,,
\label{YM--model}
\end{align}
where $\eta_{ab}$ denotes the (flat) metric of a $D$ dimensional embedding space, with  
arbitrary signature. 
The ``covariant coordinates`` $X^a$ are Hermitian matrices, resp.
operators acting on a Hilbert space $\cH$. It was shown in Ref.~\cite{Steinacker:2008ri} 
that if one considers some of the coordinates to be functions of the remaining ones
such that $X^a \sim x^a = (x^\mu,\phi^i(x^\mu))$ in the semi-classical limit,
one can interpret the $x^a$ as defining the embedding of a $2n$-dimensional submanifold 
$\cM^{2n}\hookrightarrow\R^D$ equipped with a non-trivial induced metric
\begin{align}
g_{\m\n}(x)&=\pa_\m x^a \pa_\n x^b\eta_{ab}\,,
\end{align}
via pull-back of $\eta_{ab}$. In the present case we consider this submanifold to be a four dimensional space-time $\cM^4$, and following~\cite{Steinacker:2008ri} we can interpret
\begin{align}
\co{X^\m}{X^\n}\sim i\th^{\m\n}(x)
\end{align}
as a Poisson structure on $\cM^4$. Furthermore, we assume that $\th^{\m\n}$ is non-degenerate, 
so that its inverse matrix $\th^{-1}_{\m\n}$ defines a symplectic form 
$\varTheta = \theta^{-1}_{\mu\nu} dx^\mu \wedge dx^\nu$ on $\cM^4$.

The essential point is now that the Poisson structure $\th^{\m\n}$ and the 
induced metric $g_{\m\n}$ combine to the effective metric
\begin{align}
G^{\m\n}&=e^{-\s}\th^{\m\r}\th^{\n\s}g_{\r\s}
\,, &
e^{-\s}&\equiv \frac{\sqrt{\det\th^{-1}_{\m\n}}}{\sqrt{\det G_{\r\s}}}\,.
\end{align}
It is, in fact, this effective metric $G_{\m\n}$ which is ``seen'' by matter~\cite{Steinacker:2007dq} 
(i.e. scalar fields, gauge fields, and fermions possibly up to conformal factors),
and which therefore must be interpreted in terms of gravity.
In the present work, we restrict ourselves to the special case of $G_{\m\n}=g_{\m\n}$ in 
4 dimensional space-time $\cM^4$. It is easy to see that this is equivalent
to $\th^{-1}_{\m\n}$ being (anti)self-dual, by which
in the case of Minkowski signature we mean $\star_g\syf = \pm i\syf$.
This requires that $\theta^{\mu\nu}$ is complexified, as discussed in 
\secref{sec:symplectic}.
The Yang-Mills action \eq{YM--model} then reduces in the semi-classical limit to
\begin{align}
S_{YM}&=-\Tr\co{X^a}{X^b}\co{X^c}{X^d}\eta_{ac}\eta_{bd}\,
\sim 4 \intg ,
\label{YM--model-cc}
\end{align}
which in General Relativity (GR) is interpreted as cosmological constant. 
We also recall that \eq{YM--model} leads to the following equation of motion for $\theta^{\mu\nu}$
\begin{align}
\nabla^\eta_G (e^{\sigma}\theta^{-1}_{\eta \nu})
&= G_{\rho\nu}\,\theta^{\rho \mu} e^{-\sigma}\partial_\mu \eta 
\,, &&
\eta\equiv \frac{e^\s}{4}G^{\m\n}g_{\m\n}=\Big|_{G=g}e^\s\,.
\label{theta-covar-id-text}
\end{align}
This equation holds identically for $G_{\m\n}=g_{\m\n}$ i.e. for
 self-dual $\theta^{\mu\nu}$, and is therefore not restricted to 
the model \eq{YM--model}.

\paragraph{Einstein-Hilbert action.}
In a previous paper~\cite{Blaschke:2010rg}, we have shown that the following combination of higher order terms 
in the matrix model semi-classically lead to the Einstein-Hilbert type of action:
\begin{align}
S_{\text{E-H}}
&= \Tr\left(2T^{ab}\Box X_a \Box X_b - T^{ab}\Box H_{ab}\right) 
\,\sim\, -2 \intg e^{2\s}R[g] \,,
\label{E-H-action}
\end{align}
where
\begin{align}
T^{ab}&=\inv{2}\aco{\co{X^a}{X^c}}{\co{X^b}{X_c}}
-\inv{4}\eta^{ab}\co{X^c}{X^d}\co{X_c}{X_d}
\,, \nn\\
H^{ab}&=\inv{2}\aco{\co{X^a}{X^c}}{\co{X^b}{X_c}}
\,, \nn\\
\Box Y&\equiv \co{X^a}{\co{X_a}{Y}}
\,.
\end{align}
Latin indices are pulled down with the (flat) background metric $\eta_{ab}$ (i.e. $X_a=\eta_{ab}X^b$), 
and $R[g]$ denotes the 
Ricci scalar with respect to the metric $G=g$ of the submanifold $\cM^4$. 
Such actions can be added by hand, but they will also arise upon quantization of the 
Yang-Mills matrix model \eq{YM--model}. It was argued in~\cite{Blaschke:2010rg} 
that the factor $e^{2\s}$ sets the scale and introduces the gravitational constant $G$.

Under reasonable conditions (such as global hyperbolicity), every 4-dimensional manifold
can be equipped with a self-dual (complexified) symplectic form $\syf$. Then 
the classical embedding theorems 
\cite{Clarke:1970,Friedman:1961} imply that one can realize every 4-dimensional geometry 
as semi-classical configuration in the matrix model with $g_{\mu\nu} = G_{\mu\nu}$.
In the present paper, we illustrate this general fact by providing an explicit construction of the 
most important solution: the Schwarzschild geometry. 
Subsequently, we also construct Reissner-Nordstr\"om (RN) geometry by following the same steps. 

There are several possible actions which extend \eq{E-H-action} beyond the case $g=G$ and 
which may imply different equations for $\theta^{\mu\nu}$ and for $e^\sigma$. We 
therefore restrict ourselves to 
the construction of geometries which are solutions to GR, equipped with self-dual $\theta^{\mu\nu}$. We do not 
check here in detail whether the above action \eq{E-H-action}  admits these spaces 
with self-dual $\theta^{\mu\nu}$ as solutions. Indeed additional terms in the action
should be expected, leading e.g. to a potential for $\sigma$ and possibly to deviations 
from $\theta^{\mu\nu}$ being self-dual.
The point of this paper is not to present final answers but to illustrate how geometries such as 
Schwarzschild are expected to arise within this class of matrix models.
In the same vein, we will also assume that the Yang-Mills 
resp. vacuum energy term \eq{YM--model-cc} is negligible compared with the Einstein-Hilbert action \eq{E-H-action},
thus setting the cosmological constant to zero.
There are several intriguing hints that the role of vacuum energy 
in this framework may be different than in GR \cite{Steinacker:2010rh}.

Furthermore, we only consider the semi-classical limit of the  matrix model in 
the present paper. 
Thus we will recover precisely the 
Schwarzschild geometry (resp. RN geometry), and the central singularity will be reflected by an
embedding which escapes to infinity as one approaches the center. 
Of course, the main appeal for this framework compared with other 
descriptions of gravity is the fact that it goes beyond the 
classical concepts of geometry: Space-time is not put in by hand
but \emph{emerges}, realized as {\nc} space with an effective geometry,
along with gauge fields and matter. Hence one should expect that {\nc} 
modifications become important as one approaches the singularity.
However, this requires to go beyond the semi-classical approximations
of this paper, which we will 
indicate by briefly discussing higher-order terms in 
the star product in \appref{sec:appendix-7dim-poisson}. 

Finally, we want to emphasize that
the actions under consideration are expected to arise 
upon quantization of Yang-Mills matrix models, 
such as the IKKT model \cite{Ishibashi:1996xs}. 
In particular the latter model is a promising candidate for 
a {\em quantum} theory of fundamental interactions including gravity.
Of course, much more work remains to be done in order to fully understand
this class of models.

\section{The Schwarzschild geometry}

We now show how the most important solution of General Relativity can emerge
from the class of extended matrix model action presented in the previous section: 
the Schwarzschild geometry.
We will restrict ourselves to the semi-classical limit here, however a possible way to
obtain higher-order corrections in $\theta^{\mu\nu}$ is discussed in \appref{sec:appendix-7dim-poisson}.

\subsection{Embedding of Schwarzschild geometry}

Our construction involves two steps:
\begin{enumerate}
\itemsep=0pt
\item[1)] the choice of a suitable embedding $\cM^4\subset \R^D$
such that the induced geometry on $\cM^4$ given by $g_{\mu\nu}$ is the Schwarzschild metric, and
\item[2)]  a suitable non-degenerate Poisson structure on $\cM^4$ which solves the e.o.m. $\nabla^\mu \theta^{-1}_{\mu\nu} = 0$ for self-dual symplectic form $\syf$.
\end{enumerate}
Both steps are far from unique a priori. However, the freedom is considerably reduced by 
requiring that the solution should be a ``local perturbation'' of an asymptotically flat
(or nearly flat) ``cosmological'' background. This is clear on physical grounds, having in mind
the geometry near a star in some larger cosmological context: it must be possible to 
approximately ``superimpose'' our solution, allowing e.g. for systems of stars and galaxies
in a natural way. This eliminates the well-known embeddings of the Schwarzschild geometry in the 
literature \cite{Kasner:1921,Fronsdal:1959zza,Kerner:2008nt}, 
which are highly non-trivial for large $r$ and cannot be superimposed 
in any obvious way. In fact we require that the embedding is asymptotically harmonic $\Box x^a \to 0$
for $r \to \infty$, in view of the fact that there may be terms in the matrix model which depend
on the extrinsic geometry, and which typically single out such harmonic 
embeddings\footnote{This can hold only asymptotically, since Ricci-flat geometries 
can in general \emph{not} be embedded harmonically \cite{Nielsen:1987}.}.

Furthermore, we insist that $\theta^{\mu\nu}$ is non-degenerate, and 
$\th^{\mu\nu} \to \const \neq 0$ as $r \to \infty$. This is again motivated 
by the requirement that physics at large distances should not be affected by a localized mass.
In particular, $e^{\sigma}$ defines essentially the scale of non-commutativity, and certainly enters in some way 
e.g. the physics of elementary particles (In fact, $e^\sigma$ determines the strength of the gauge coupling
in the matrix model \cite{Steinacker:2007dq,Steinacker:2008ri}). 
Therefore, $e^{\sigma}$ should be asymptotically constant and non-vanishing. This is an important difference to 
previous proposals for a {\nc} Schwarzschild geometry 
(see in particular \cite{Schupp:2009pt,Ohl:2008tw} and references therein), 
where the Poisson structure is degenerate and/or not asymptotically constant. 
Hence $\theta^{\mu\nu}$ will be viewed as some cosmological background field which is locally perturbed by
a mass. Recall that such a background is essentially invisible, since there are no 
fields in the matrix model which are charged
under the corresponding $U(1)$. It enters the effective actions only through the 
gravitational metric $G_{\mu\nu}$.

Note that these boundary conditions for $\theta^{\mu\nu}$ are not in conflict with the 
idea that $\theta^{\mu\nu}$ may be locally fluctuating and 
should perhaps be averaged over or integrated out. We will discuss this possibility further below.

In order to obtain an appropriate embedding, 
keeping in mind the conditions we have just discussed, we consider Eddington-Finkelstein coordinates and define:
\begin{align}
t &= t_S + (r^* - r)\,,  
&
r^* &= r + r_c\ln\left|\frac{r}{r_c} - 1\right|
\,,
\end{align}
where $t_S$ denotes the usual Schwarzschild time, $r_c$ is the horizon of the Schwarzschild black hole and $r^*$ is 
the well-known tortoise coordinate. The metric in Eddington-Finkelstein coordinates $\{t,r,\vth,\vph\}$ is
\begin{align}
 ds^2 &= - \left( 1-\frac{r_c}{r} \right) dt^2 + \frac{2 r_c}{r}
    dt dr + \left( 1 + \frac{r_c}{r} \right) dr^2 + r^2 d\Omega^2 
\label{eq:eddington-finkelstein-metric}
\end{align}
which is asymptotically flat for large $r$, and
manifestly regular at the horizon $r_c$. 
Thus, we only need to find a good embedding of
this metric. To reproduce the mixed term, consider first
\bea
\phi_1+i\phi_2 = h(r) e^{i(\omega t + g(r))}
\eea
which satisfies
\begin{align}
\pa_t\phi^i \pa_t \phi^i &= \w^2 h^2 \,, \nn\\
\pa_t\phi^i \pa_r \phi^i &= \w g' h^2 \,, \nn\\
\pa_r\phi^i \pa_r \phi^i &= g'^2 h^2 + h'^2
\,. 
\label{eq:embedding-zw1}
\end{align}
So we demand
\begin{align}
\w g' h^2 &= \frac{r_c}{r}
=
\w^2 h^2 
\end{align}
which is satisfied for 
\begin{align}
h(r) &= \inv{\w}\sqrt{\frac{r_c}{r}}
\,, &
g(r) &= \w r\,.
\end{align}
Furthermore, since $g'^2 h^2=\omega g' h^2=\frac{r_c}{r}$, we need to cancel the $h'^2$ term in \eqnref{eq:embedding-zw1} above. 
Hence, we need another coordinate
\be
\phi_3 = h(r),
\ee
with time-like embedding. So we have
\begin{align}
 \phi_1+i\phi_2&=\phi_3e^{i\omega (t + r)}\,,\nn\\
\phi_3&=\inv{\w}\sqrt{\frac{r_c}{r}}\,,
\label{eq:embedding-fcts}
\end{align}
and the embedding of $\cM^4 \subset \R^7$ is given by
\begin{align}
x^a&=\left(\begin{array}{c}
 t  \\
 r\cos\vph\sin\vth \\
 r\sin\vph\sin\vth \\
 r\cos\vth \\
 \inv{\w}\sqrt{\frac{r_c}{r}}\cos\left(\w(t+r)\right) \\[0.3em]
 \inv{\w}\sqrt{\frac{r_c}{r}}\sin\left(\w(t+r)\right) \\[0.3em]
 \inv{\w}\sqrt{\frac{r_c}{r}}
\end{array}\right)\,,
\label{eq:embedding}
\end{align}
(i.e. we consider $D=7$ in this example). Together with the background metric
\begin{align}
\eta_{ab}&=\diag(-,+,+,+,+,+,-)\,,
\end{align}
this induces precisely the Eddington-Finkelstein
metric \eqref{eq:eddington-finkelstein-metric} above.

\begin{figure}[!ht]
\centering
\vspace{0.5cm}
\includegraphics[scale=0.9]{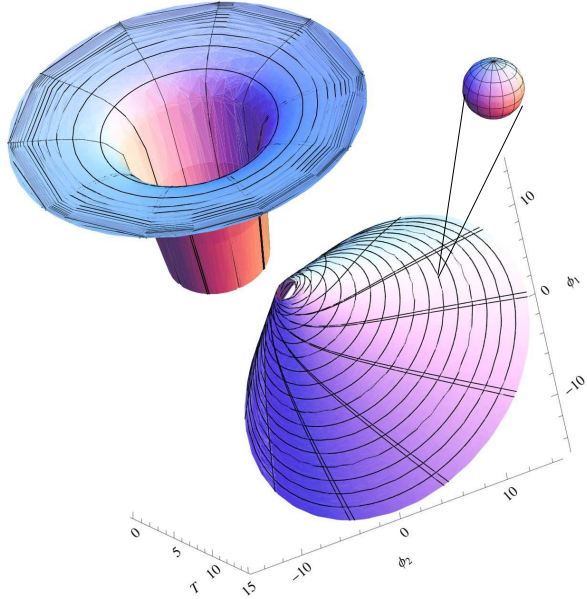}
\caption{Embedded Schwarzschild black hole. $\ $ \textit{ On the top, a schematic view of the outer region of the 
Schwarzschild black hole is shown. After passing through the horizon $r=r_c$, the extra dimensions $\phi_i$ ``blow up'' 
in a cone-like manner. As indicated in the lower half of this figure, every point of the cone is in fact a sphere whose 
radius $r$ becomes smaller towards the bottom of the cone (i.e. $T\propto1/\sqrt{r}$). The twisted vertical lines drawn 
in the cone are lines of equal time $t$.}}
\label{fig:black-hole}
\end{figure}

Before we proceed to determine the symplectic form, let us take a closer look at the properties of this embedding: 
First, notice that the $\w$ (appearing in the $\phi_i$) does not enter the effective four dimensional metric, 
i.e. it is ``hidden'' in the three extra dimensions. Furthermore, we must emphazise that $\phi_3$ is an additional 
\emph{time-like} direction. Asymptotically, i.e. for $r\to\infty$, \eqnref{eq:embedding} describes flat four dimensional Minkowski 
space where the extra dimensions $\phi_i\sim\inv{r}$ become infinitesimally small. On the other hand, when one approaches 
the singularity of the Schwarzschild black hole at $r=0$, these extra dimensions blow up and 
become arbitrarily large. 
In particular, note that then $\phi_3$ should be interpreted as asymptotic time (which is unbounded), 
i.e. 
\be
T:= \phi_3 \,, \qquad
r = r_c \, \frac 1{\w^2 T^2}
\,,
\ee
so that
\bea
x^a = \left(\begin{array}{cc}
x^0 &  \\
x^i & \\
\phi_1+ i \phi_2 & \\
\phi_3
\end{array}\right) 
= \left(\begin{array}{c}
 t,  \\
r_c \, \frac 1{\omega^2 T^2}\, 
(\sin\vth\sin\vph,\sin\vth\cos\vph,\cos\vth) \\
T \,e^{i \w (t+r_c \, \inv{\w^2 T^2})} \\
T
\end{array}\right)
\label{finkelstein-embedding-int}
\eea
for large $T$. This is a helicoid-like (cone-like) geometry in $\{T,t\}$ with (increasing) radius $T$ and $t$ 
playing the role of the angle variable, times a contracting sphere of radius $\inv{T^2}$. 
The geometry of the submanifold $\cM^{1,3} \subset \R^{2,5}$ is completely regular, and the central singularity is 
reflected by an embedding which escapes to infinity. Near this singularity, the geometry 
is effectively 2-dimensional.
An illustration is given by Figure~\ref{fig:black-hole}. 

Of course, quantum effects will play a major role near $r=0$. This implies that the semi-classical approximation we are currently considering 
will break down in the vicinity of that region. We expect that these {\nc} effects 
will regularize the would-be singularity. 
For example, the contracting sphere of radius $\inv{T^2}$ may become fuzzy 
\cite{Madore:1991bw}, so that for 
large $T$ the present geometry could become 
effectively 2-dimensional with an extra-dimensional fuzzy sphere.

In order to understand the meaning of the extra dimensions $\phi_i$ at large distances $r\to\infty$ 
from the Schwarzschild black hole, 
it is instructive to consider the following modification resp. higher-dimensional extension of the
Schwarzschild geometry.
Consider the 6-dimensional space $\R^4 \times AdS^2 \subset \R^7$ defined by
\begin{align}
\phi_1^2+\phi_2^2-\phi_3^2=R^2\,.
\end{align}
Here the $\phi_i$ describe an $AdS^2$ space embedded in $\R^3$, which can be parametrized as
\begin{align}
\phi_1+i \phi_2 &= \sqrt{\phi_3^2+ R^2}\, e^{i\w u}\,,\nn\\
\phi_3 &= \phi_3\,.
\end{align}
The Schwarzschild manifold described above is then recovered by setting
\begin{align}
u = t+r\,, \qquad \phi_3 = \inv{\w} \sqrt{\frac{r_c}{r}}\,, \quad \text{and } R=0\, ,
\end{align}
while $R \neq 0$ corresponds to a modification of the Schwarzschild geometry.
The length element of $AdS^2\subset \R^3$ is given by
\begin{align}
ds^2 = d\phi_1^2+d\phi_2^2-d\phi_3^2 = \w^2 \left(\phi_3^2+ R^2\right) du^2 - \frac{R^2}{\phi_3^2+R^2} d \phi_3^2 \,.
\label{metric-AdS}
\end{align}
Note that there is no contribution from the time-like coordinate $d\phi_3^2$ for $R=0$ since it is embedded in a null direction. 
The metric on $AdS^2$ then becomes degenerate and space-like, with very small radius as $r\to\infty$. 
Hence, this extra $AdS^2$ can be interpreted as physical extra dimension which naturally 
becomes ``invisible'' for large $r$, i.e. far away from the Schwarzschild black hole.
The point is that such an $AdS^2$ could arise naturally in matrix models similar to fuzzy spheres,
and may play an interesting physical role, cf. 
\cite{Ho:2000fy,Alekseev:2000fd,Iso:2001mg,Aschieri:2006uw}.

\subsection{Symplectic form}
\label{sec:symplectic}
As mentioned in the introduction we consider the simple class of self-dual geometries where the 
effective metric $G_{\m\n}$ equals the induced metric $g_{\m\n}$. 
Hence we need to find an (anti)self-dual symplectic form $\syf$ so that
\begin{align}
G^{\m\n}&=e^\s\th^{\m\r}\th^{\n\s}g_{\r\s}\nn\\
&=g^{\m\n}\,.
\label{eq:Gisg}
\end{align}
At this point, we recall that
\be
\cJ^{\eta}_\g = e^{-\sigma/2}\, \theta^{\eta\g'} g_{\g' \g} 
\ee
satisfies
\be
\cJ^2 = -1 \quad\Leftrightarrow \quad \star\syf = \pm i \syf\, 
\quad\Leftrightarrow \quad g_{\mu\nu} = G_{\mu\nu}\, ,
\ee
so that we are dealing with an almost complex manifold. Moreover,
the symplectic structure is necessarily complexified in a way which is determined
by $\cJ^2 = -1$. Thus the last relation specifies\footnote{In the case of general geometries $g_{\mu\nu} \neq G_{\mu\nu}$ 
this is replaced by a quartic relation for $\cJ$  \cite{Steinacker:2008ya}.} the ``real form'' of $\theta^{\mu\nu}$.

Furthermore, we require $\syf= \theta^{-1}_{\mu\nu}dx^\mu\wedge dx^\nu$ to lead to an asymptotically constant $e^{-\s}$ since, 
as mentioned previously, we would like to describe everything as a local perturbation of flat Moyal space. 
To be more specific, we demand
\begin{align}
\lim\limits_{r\to\infty}e^{-\s}&=\const\neq0\,.
\label{eq:esig-asym}
\end{align}
In order to find such a symplectic form $\syf$, we consider the following: The Schwarzschild metric has two Killing vector fields $V_{ts}=\pa_{t_S}$ and $V_\vph=\pa_\vph$. 
Hence, in Schwarzschild coordinates this leads to the ansatz
\begin{align}
\syf &= i \syf_E + \syf_B 
= i E \wedge dt_S + \syf_B, \nn\\
E &= i_{V_{ts}} \syf_E 
= E_r dr + E_\vth d\vth + E_\vph d\vph, \nn\\
\syf_B &=  B_r d\vth\wedge d\vph + B_\vth dr\wedge d\vph
+ B_\vph d\vth \wedge d r\nn\\
&= \star \syf_E\,,
\label{eq:syf-ansatz}
\end{align}
which implements self-duality, i.e.
\begin{align}
\star\syf= i \syf
\,,
\end{align}
supplemented by the conditions
\begin{align}
\cL_{V_{ts}}\syf&=0\,,\nn\\
\cL_{V_\vph}\syf&=0\,.
\end{align}
A solution which satisfies the required asymptotics \eqref{eq:esig-asym} is then given by
\begin{align}
E &= \cst\left( \cos\vth dr -r(1-\frac{r_c}{r})\sin\vth d\vth\right)
 = d(f(r)\cos\vth)\,,\nn\\
B &= \cst\left(r^2\sin\vth \cos\vth d\vth + r \sin^2\vth dr\right)
= \frac{\cst}{2} d(r^2\sin^2\vth)\,,\nn\\
\syf &= i E\wedge dt_S+B\wedge d\vph\,,\nn\\
&\text{with }\ f(r)=\cst r(1-\frac{r_c}{r}), \qquad f' = \cst=\const\,,
\label{eq:symplectic-solution}
\end{align}
from which one finds
\begin{align}
e^{-\s}&=\cst^2\left(1-\frac{r_c}{r}\sin^2\vth\right)\,.
\label{eq:esig-solution}
\end{align}
Details of the computation are given in \appref{sec:appendix-symplectic}. 
This can be interpreted as a (complexified) electromagnetic field with 
asymptotically constant fields $E,B$ pointing in the $z$ direction,
and $e^{-\s}$ is indeed asymptotically constant.
Other solutions are of course obtained by acting with the rotation group on the 
asymptotic $E$ resp. $B$ field.

Note  in particular that we have obtained metric-compatible Darboux coordinates 
(resp. Hamiltonian reduction) $x_D^\m=\{H_{ts},t_S,H_\vph,\vph\}$ corresponding 
to $V_{ts},\, V_\vph$ where the symplectic form is constant:
\begin{align}
\syf &= i\cst d H_{ts} \wedge dt_S + \cst d H_\vph \wedge d\vph\,, \nn\\
&=\cst d\left(iH_{ts}dt_S+H_\vph d\vph\right)\,,\nn\\
H_{ts} &= r\cos\vth(1-\frac{r_c}{r})\,,
\qquad H_\vph = \inv{2} r^2\sin^2\vth\,.
\label{eq:darboux}
\end{align}
The Schwarzschild metric in Darboux coordinates reads
\begin{align}
ds^2&=-\left(1-\frac{r_c}{r}\right)dt_S^2+\frac{e^\bs}{\left(1-\frac{r_c}{r}\right)}dH_{ts}^2+r^2\sin^2\vth d\vph^2+\frac{e^\bs}{r^2\sin^2\vth}dH_\vph^2\,,
\label{eq:ss-metric-darboux}
\end{align}
with $e^\bs=c_1^2e^\s=\left(1-\frac{r_c}{r}\sin^2\vth\right)^{-1}$. Notice that no $dH_{ts}dH_\vph$-term appears, 
i.e. the two Darboux blocks do not mix. 
The relations to the Killing vector fields are:
\begin{align}
E &= \cst d H_{ts}  = \cst E_\m dx^\m = i_{V_{ts}} \syf \,,   && E_\m = V_{ts}^\n \th^{-1}_{\n\m}\,,   \nn\\
B &= \cst d H_\vph = \cst B_\m dx^\m = i_{V_\vph} \syf \,,  && B_\m = V_\vph^\n \th^{-1}_{\n\m}\,,
\end{align}
(cf. \eqnref{eq:symplectic-solution} above). 
In order to obtain the Poisson brackets between the Cartesian matrix coordinates, 
we will transform $t_S$ to Eddington-Finkelstein time $t$ and invert it so as to derive $\th^{\m\n}$. Subsequently, we will extend the $\th$-matrix to the seven dimensional embedding space of \eqnref{eq:embedding} 
as that will provide us with the leading order commutator relations between the coordinates, i.e. $\co{X^a}{X^b}\sim i\th^{ab}$. 
As shown in \appref{sec:appendix-7dim-poisson}, this leads to the following semi-classical commutation relations 
for the 7-dimensional coordinates $X^a\sim x^a=\{t,x,y,z,\phi_1,\phi_2,\phi_3\}$:
\begin{align}\label{eq:theta7-dir_c}
\th^{ab}\!\!&= \sth e^\bs \!\left(\!\!\!\begin{array}{ccccccc}
 0 & -\frac{r_cy}{r^2} & \frac{r_cx}{r^2} & -i & \frac{i z}{r}f^+_{12}(0) & \frac{i z}{r}f^-_{21}(0) & \frac{i z\phi_3}{2r^2} \\
 \frac{r_cy}{r^2} & 0 & e^{-\bs} & -\frac{r_cyz}{r^3}   & -\frac{y}{r}f^+_{12}(r_c) & -\frac{y}{r}f^-_{21}(r_c) & -\frac{y\g \phi_3}{2r^2}  \\
 -\frac{r_cx}{r^2} & -e^{-\bs} & 0 & \frac{r_cxz}{r^3} & \frac{x}{r}f^+_{12}(r_c) & \frac{x}{r}f^-_{21}(r_c) & \frac{x\g \phi_3}{2r^2} \\
 i & \frac{r_cyz}{r^3} & -\frac{r_cxz}{r^3} & 0 & -i\w\phi_2 & i\w\phi_1 & 0 \\
-\frac{iz}{r}f^+_{12}(0) & \frac{y}{r}f^+_{12}(r_c) & -\frac{x}{r}f^+_{12}(r_c) & i\w\phi_2 & 0 & -\frac{i\w z\phi_3^2}{2r^2} & -\frac{i\w z\phi_3\phi_2}{2r^2} \\
-\frac{i z}{r}f^-_{21}(0) & \frac{y}{r}f^-_{21}(r_c) & -\frac{x}{r}f^-_{21}(r_c) & -i \w\phi_1 & \frac{i\w z\phi_3^2}{2r^2} & 0 & \frac{i\w z\phi_3\phi_1}{2r^2} \\
 -\frac{i z\phi_3}{2r^2} & \frac{y\g \phi_3}{2r^2} & -\frac{x\g \phi_3}{2r^2} & 0 & \frac{i\w z\phi_3\phi_2}{2r^2} & -\frac{i\w z\phi_3\phi_1}{2r^2} & 0
\end{array}\!\!\!\right)\!,
\end{align}
with
\begin{align}
f^{\pm}_{ij}(r_c)&=\left(\frac{\g}{2r}\phi_i\pm\w\phi_j\right)\,,\nonumber\\
\g&=\left(1-\frac{r_c}{r}\right)\,,\nonumber\\
e^{-\bs}&=\frac{e^{-\s}}{\cst^2}=\sth^2 e^{-\s}\,.
\end{align}
This defines a Poisson structure on $\cM^4$, but it could also be viewed as a Poisson structure
on the 6-dimensional space defined by $\phi_1^2 + \phi_2^2 = \phi_3^2$ which admits $\cM^4$ as symplectic leaf.
As a consistency check, the interested reader may verify that relation \eqref{eq:Gisg} is indeed fulfilled 
(on the 4-dimensional submanifold $\cM^4$), and that the Jacobi identity 
holds as well.

\subsection{Star product}
So far, we have worked only in the semi-classical limit. In order to see some effects of the space-time quantization, 
we may for instance compute the next-to-leading order commutation relations. For this purpose, recall the Darboux coordinates $x_D^\m=\{t_S, H_{ts},\vph, H_\vph \}$ we derived in \eqref{eq:darboux}. Since in these coordinates the Poisson structure $\th^{\m\n}$ (of the 4 dim. submanifold $\cM^4$) is constant, we can easily define a Moyal-type star product~\cite{Douglas:2001ba,Szabo:2001kg} as
\begin{align}
(g\star h)(x_D) &= g(x_D) e^{-\frac{i}{2}\left(\lpa_\m\th_{D}^{\m\n}\rpa_\n\right)} h(x_D) \,,
\end{align}
with
\begin{align}
\th_{D}^{\m\n}&=\sth\left(\begin{array}{cccc}
0 & i & 0 & 0\\
-i & 0 & 0 & 0\\
0 & 0 & 0 & 1\\
0 & 0 & -1 & 0
\end{array}\right)\,,
\end{align}
where $\sth=1/\cst\ll1$ denotes the expansion parameter. 
In order to derive a star product in terms of the Cartesian coordinates $x^\m=\{t,x,y,z\}$, all we need is the coordinate transformation \eqref{eq:app-coordinate-trafo} of \appref{sec:appendix-7dim-poisson} leading to
\begin{align}
(g\star h)(x) &= g(x) \exp\Bigg[\frac{i\sth}{2}\Bigg(\left(\lpa_t\frac{ir_cz e^\bs}{r(r-r_c)}+\lpa_zie^\bs\right)\wedge\rpa_t\nonumber\\*
&\quad+\left(\left(\lpa_t-\lpa_z\frac{z}{r}\right)\frac{r_c e^\bs}{r^2}+\left(\lpa_xx+\lpa_yy\right)\inv{x^2+y^2}\right)\wedge\left(x\rpa_y-y\rpa_x\right)\Bigg)\Bigg] h(x) \,,
\end{align}
where the wedge stands for ``antisymmetrized'', and when considering the expansion one must take care with the sequence of operators and the side they act on (left or right). 
One can then compute next-to-leading order contributions to the commutation relations \eqref{eq:theta7-dir_c}. 
Some of the relations can be computed to all orders\footnote{It is also interesting to note, that the quantities $\{z,\,\phi_3,\,H_{\vth_2}\}$, where 
\begin{align*}
H_{\vph}&=\tinv{2}x_+x_-=\tinv{2}(x^2+y^2)=\tinv{4}\staraco{x_+}{x_-}
\,, &
x_\pm&=x\pm i y\,,
\end{align*}
commute with each other to all orders in $\sth$.}, i.e.
\begin{align}
\starco{t}{z}&=\sth e^\bs
\,, &
\starco{x}{y}&= i\sth\,,\nonumber\\
\starco{t}{\phi_3}&= -\sth e^\bs\frac{z\phi_3}{2r^2}
\,, &
\starco{z}{\phi_3}&= 0\,,
\label{eq:exact-commutators}
\end{align}
while the others receive corrections --- see Eqns. \eqref{eq:nlo-co-rel-txyz}-\eqref{eq:nlo-co-rel-phi} in \appref{sec:appendix-7dim-poisson} for the full expressions.
Hence, also the embedding constraint $\phi_1^2+\phi_2^2=\phi_3^2$ is modified under the star product, i.e. we have
\begin{align}
\inv{2}\staraco{\phi_1+ i\phi_2}{\phi_1- i\phi_2}&=\phi_1\star\phi_1+\phi_2\star\phi_2\nonumber\\
&=\phi_3^2+\sth^2\phi_3^2\frac{\w^2 e^{2\bs}}{8r^2}\left(1-3\frac{ e^\bs z^2}{r^2}\right)+\mO{\sth^4}\,,
\end{align}
while $\phi_3\star\phi_3=\phi_3^2$ to all orders. 
This could be interpreted as {\nc} correction to the embedding geometry.

\section{The Reissner-Nordstr{\"o}m geometry}

In this section, we continue by presenting the semi-classical quantization of another geometry: 
the Reissner-Nordstr{\"o}m (RN) geometry.

\subsection{Embedding of the geometry}
We start by considering the usual RN metric in spherical coordinates $x^\m=\{t,r,\vth,\vph\}$:
\begin{align}
ds^2=-\left(1-\frac{2m}{r}+\frac{q^2}{r^2}\right)d\Tt^2+\inv{\left(1-\frac{2m}{r}+\frac{q^2}{r^2}\right)}dr^2+r^2d\W
\,, 
\end{align}
where $m$ denotes the mass and $q$ is the charge of the black hole. 
This geometry has two concentric horizons at
\begin{align}
r_h&=\left(m\pm\sqrt{m^2-q^2}\right)\,,
\label{eq:RN-horizons}
\end{align}
and in the following, we assume that $q^2<m^2$. In order to transform this metric into coordinates which are similar to Eddington-Finkelstein, we consider radial null geodesics. These are given by
\begin{align}
0&=-\left(1-\frac{2m}{r}+\frac{q^2}{r^2}\right)\left(d\Tt^2-\inv{\left(1-\frac{2m}{r}- \frac{q^2}{r^2}\right)^2}dr^2\right)\nonumber\\
&\equiv -\left(1-\frac{2m}{r}+\frac{q^2}{r^2}\right)\left(d\Tt^2-(dr^*)^2\right)\,,
\end{align}
defining the tortoise-like coordinate $r^*$. The in and outgoing geodesics are $V=\Tt+r^*$ and $U=\Tt-r^*$. Explicitly, we have
\begin{align}
r^*&=r+m\ln\left|r^2-2mr+q^2\right|+\frac{2m^2-q^2}{2\sqrt{m^2-q^2}}\ln\left|\frac{\sqrt{m^2-q^2}-(r-m)}{\sqrt{m^2-q^2}+(r-m)}\right|\,.
\end{align}
As in the Schwarzschild case, we use this coordinate to shift the time-coordinate according to
\begin{align}
t=\Tt+(r^*-r)\,,
\end{align}
and arrive at the transformed RN metric
\begin{align}
ds^2=-\left(1-\frac{2m}{r}+\frac{q^2}{r^2}\right)dt^2+2\left(\frac{2m}{r}-\frac{q^2}{r^2}\right)dtdr+\left(1+\frac{2m}{r}-\frac{q^2}{r^2}\right)dr^2+r^2d\W\,.
\label{eq:RN-metric-smooth}
\end{align}
Observe, that the metric \eqref{eq:RN-metric-smooth} has exactly the same form as the Eddington-Finkelstein metric \eqref{eq:eddington-finkelstein-metric} of Schwarzschild geometry, but with the replacement
\begin{align}
\frac{r_c}{r}\to\frac{2m}{r}-\frac{q^2}{r^2}\,.
\end{align}
Hence, motivated by the Schwarzschild geometry case, we can use the 10-dimensional embedding $\cM^{1,3}\hookrightarrow\R^{4,6}$ with the additional coordinates $\phi_i$ given by
\begin{align}
 \phi_1+i\phi_2&=\phi_3e^{i\w (t + r)}\,,\qquad
&\phi_3&=\inv{\w}\sqrt{\frac{2m}{r}}\,,\nonumber\\
\phi_4+i\phi_5&=\phi_6e^{i\w (t+r)}\,,\qquad
&\phi_6&=\frac{q}{\w r}
\,.
\label{eq:RN-embedding-fcts}
\end{align}
Note that $\phi_3$, $\phi_4$ and $\phi_5$ are \emph{time-like} coordinates, i.e. we consider the background metric
\begin{align}
\eta_{ab}&=\diag(-,+,+,+,+,+,-,-,-,+)\,.
\end{align}
Like in the previous case, $\w$ does not enter the induced metric \eqref{eq:RN-metric-smooth}, but is hidden in the extra dimensions $\phi_i$. For $r\to\infty$, the $\phi_i$ become infinitesimally small and hence asymptotically, the four dimensional subspace becomes flat Minkowski space-time.

\subsection{Symplectic form}
A self-dual symplectic form $\syf$ can be computed in the same way as in the Schwarzschild case leading to metric compatible Darboux coordinates $x_D^\m=\{H_{\Tt},\Tt,H_\vph,\vph\}$ with
\begin{align}
H_{\Tt}&=r\g \cos\vth=z\g\,, \qquad\qquad
H_\vph=\a\frac{r^2}{2}\sin^2\vth=\a\frac{x^2+y^2}{2}\,,\nonumber\\
\syf&=idH_{\Tt}\wedge d\Tt+dH_\vph\wedge\vph\,,\nonumber\\
e^{-\bs}&=\g\sin^2\vth+\a^2\cos^2\vth 
=\g\left(1-\frac{q^2z^2}{r^4}\right)+\a\eta\frac{z^2}{r^2}\,,\nonumber\\
\g&= \left(1-\frac{2m}{r}+\frac{q^2}{r^2}\right)\,, \qquad
\a=\left(1-\frac{q^2}{r^2}\right)\,,\qquad
\eta= 2\left(\frac{m}{r}-\frac{q^2}{r^2}\right)
\,, 
\label{eq:RN-Darboux}
\end{align}
and the RN metric in Darboux coordinates reads
\begin{align}
ds^2=-\g d\Tt^2+\frac{e^\bs}{\g}dH_{\Tt}^2+r^2\sin^2\vth d\vph^2+\frac{e^\bs}{r^2\sin^2\vth}dH_\vph^2
\,,
\end{align}
a form similar to the according Schwarzschild metric \eqref{eq:ss-metric-darboux}. 
In the limit $q\to0$ these expressions reduce to those in the Schwarzschild case. Furthermore, one can easily check 
that $\star\syf=i\syf$ and $G_{\m\n}=g_{\m\n}$.

\subsection{Star product}
A Moyal type star product can easily be defined in Darboux coordinates as
\begin{align}
(g\star h)(x_D) &= g(x_D) e^{-\frac{i}{2}\left(\lpa_\m\th_{D}^{\m\n}\rpa_\n\right)} h(x_D) \,,
\end{align}
with
\begin{align}
\th_{D}^{\m\n}&=\e\left(\begin{array}{cccc}
0 & i & 0 & 0\\
-i & 0 & 0 & 0\\
0 & 0 & 0 & 1\\
0 & 0 & -1 & 0
\end{array}\right)
\,, \qquad \e\in\R \,.
\end{align}
Transforming these Darboux coordinates back to the Cartesian ones, where
\begin{align}
t=\Tt+(r^*-r)\,,\qquad r^2=x^2+y^2+z^2\,,
\end{align}
we eventually find
\begin{subequations}
\begin{align}
\pa_t=\pa_{\Tt}\,,\qquad \pa_\vph=-y\pa_x+x\pa_y\,,
\label{eq:RN-pat-pavph}
\end{align}
and
\begin{align}
\pa_{H_{\Tt}}&=e^\bs\left[\a\frac{z}{r}\left(\inv{\g}-1\right)\pa_t+\left(1-\frac{q^2z^2}{r^4}\right)\pa_z-\frac{q^2z}{r^4}\left(x\pa_x+y\pa_y\right)\right]\,,\nonumber\\
\pa_{H_\vph}&=\frac{e^\bs}{r}\left[\left(\g-1\right)\pa_t-\eta\frac{z}{r}\pa_z\right]+\frac{e^{\bs-\varsigma}}{x^2+y^2}\left(x\pa_x+y\pa_y\right)\,,\nonumber\\
e^{-\varsigma}&=\left(\g+\eta\frac{z^2}{r^2}\right)
\,,
\label{eq:RN-paHt-paHvph}
\end{align}
\end{subequations}
(cf. the abbreviations defined in \eqnref{eq:RN-Darboux}). The star product in Cartesian coordinates hence reads
\begin{align}
(g\star h)(x) &= g(x) \exp\Bigg[\frac{i\e}{2}\left(i\lpa_{H_{\Tt}}\wedge\rpa_t+\lpa_{H_\vph}\wedge\rpa_\vph\right)\Bigg] h(x) \,,
\end{align}
with \eqnref{eq:RN-pat-pavph} and \eqnref{eq:RN-paHt-paHvph}, where once more the wedge stands for ``antisymmetrized'', and when considering the expansion one must take care with the sequence of operators and the side they act on (left or right). The first order results for the star commutators between the 10-dimensional embedding coordinates are given by Eqns.~\eqref{eq:starcom-lo}:
\begin{subequations}\label{eq:starcom-lo}
\begin{align}
-i\starco{x^\m}{x^\n}\approx\th^{\m\n}&=\e e^\bs \!\!\left(\!\!\!\begin{array}{cccc}
 0 & \frac{-(1-\g)y}{r}+\frac{iq^2xz}{r^4} & \frac{(1-\g)x}{r}+\frac{iq^2yz}{r^4} & -i\beta \\
 \frac{(1-\g)y}{r} & 0 & e^{-\varsigma} & \frac{-yz\eta}{r^2} \\
 \frac{-(1-\g)x}{r} & -e^{-\varsigma} & 0 & \frac{xz\eta}{r^2} \\
 i\beta  & \frac{yz\eta}{r^2} & \frac{-xz\eta}{r^2} & 0 
\end{array}\!\!\!\right)\!,
\end{align}
\begin{align}
&-i\starco{\phi_i}{x^\m}\approx \e e^\bs \!\!\left(\!\!\!\begin{array}{cccc}
 \frac{-iz\alpha f^+_{12}\left(\inv{2}\right)}{r} & \frac{yf^+_{12}\left(\frac{\g}{2}\right)}{r}-\frac{iq^2xz\w\phi_2}{r^4} & \frac{-xf^+_{12}\left(\frac{\g}{2}\right)}{r}-\frac{iq^2yz\w\phi_2}{r^4} & i\w\phi_2\beta \\
\frac{-i z\alpha f^-_{21}\left(\inv{2}\right)}{r} & \frac{yf^-_{21}\left(\frac{\g}{2}\right)}{r}+\frac{iq^2xz\w\phi_1}{r^4} & \frac{-xf^-_{21}\left(\frac{\g}{2}\right)}{r}+\frac{iq^2yz\w\phi_1}{r^4} & -i \w\phi_1\beta \\
 \frac{-i z\phi_3\alpha}{2r^2} & \frac{y\g \phi_3}{2r^2} & \frac{-x\g \phi_3}{2r^2} & 0 \\
 \frac{-iz\alpha f^+_{45}(1)}{r} & \frac{yf^+_{45}(\g)}{r}-\frac{iq^2xz\w\phi_5}{r^4} & \frac{-xf^+_{45}(\g)}{r}-\frac{iq^2yz\w\phi_5}{r^4} & i\w\phi_5\beta \\
 \frac{-iz\alpha f^-_{54}(1)}{r} & \frac{yf^-_{54}(\g)}{r}+\frac{iq^2xz\w\phi_4}{r^4} & \frac{-xf^-_{54}(\g)}{r}+\frac{iq^2yz\w\phi_4}{r^4} & -i\w\phi_4\beta \\
 \frac{-iz\phi_6\alpha}{r^2} & \frac{y\g\phi_6}{r^2} & \frac{-x\g\phi_6}{r^2} & 0 
\end{array}\!\!\!\right)\!,
\end{align}
\begin{align}
&-i\starco{\phi_i}{\phi_j}\approx \e e^\bs \!\!\left(\!\!\!\begin{array}{cccccc}
 0 & \frac{-i\w z\phi_3^2\alpha}{2r^2} & \frac{-i\w z\phi_3\phi_2\alpha}{2r^2} & \frac{-i\w z\phi_1\phi_5\alpha}{2r^2} & \frac{-i\w z\alpha g_{\phi}}{2r^2} & \frac{-i\w z\phi_3\phi_5\alpha}{r^2} \\
 \frac{i\w z\phi_3^2\alpha}{2r^2} & 0 & \frac{i\w z\phi_3\phi_1\alpha}{2r^2} & \frac{-i\w z\alpha g_{\phi}}{2r^2} & \frac{i\w z\phi_2\phi_4\alpha}{2r^2} & \frac{i\w z\phi_3\phi_4\alpha}{r^2} \\
 \frac{i\w z\phi_3\phi_2\alpha}{2r^2} & \frac{-i\w z\phi_3\phi_1\alpha}{2r^2} & 0 & \frac{i\w z\phi_3\phi_5\alpha}{2r^2} & -\frac{i\w z\phi_3\phi_4\alpha}{2r^2} & 0 \\
 \frac{i\w z\phi_1\phi_5\alpha}{2r^2} & \frac{i\w z\alpha g_\phi}{2r^2} & \frac{-i\w z\phi_3\phi_5\alpha}{2r^2} & 0 & \frac{-i\w z\phi_6^2\alpha}{r^2} & \frac{-i\w z\phi_5\phi_6\alpha}{r^2} \\
 \frac{i\w z\alpha g_\phi}{2r^2} & \frac{-i\w z\phi_2\phi_4\alpha}{2r^2} & \frac{i\w z\phi_3\phi_4\alpha}{2r^2} & \frac{i\w z\phi_6^2\alpha}{r^2} & 0 & \frac{i\w z\phi_4\phi_6\alpha}{r^2}  \\
 \frac{i\w z\phi_3\phi_5\alpha}{r^2} & \frac{-i\w z\phi_3\phi_4\alpha}{r^2} & 0 & \frac{i\w z\phi_5\phi_6\alpha}{r^2} & \frac{-i\w z\phi_4\phi_6\alpha}{r^2} & 0 
\end{array}\!\!\!\right)\!,
\end{align}
with
\begin{align}
f^{\pm}_{ij}(Y)&=\left(\frac{Y}{r}\phi_i\pm\w\phi_j\right)
\,, \qquad
\beta =\left(1-\frac{q^2z^2}{r^4}\right)\,,\nonumber\\
g_{\phi}&=\left(\phi_3\phi_6+\phi_1\phi_5\right)=\left(\phi_3\phi_6+\phi_2\phi_4\right)\,,
\end{align}
\end{subequations}
and the abbreviations defined in \eqnref{eq:RN-Darboux}. 
Although some of these commutators are exact to all orders, i.e.
\begin{align}
\starco{z}{\phi_3}&=\starco{z}{\phi_6}=\starco{\phi_3}{\phi_6}=0
\,, &
\starco{t}{z}&= -i\e e^\s\beta\,,\nonumber\\
\starco{t}{\phi_3}&= i\e e^\bs\frac{z\phi_3\alpha}{2r^2}
\,, &
\starco{t}{\phi_6}&= i\e e^\bs\frac{z\phi_6\alpha}{r^2}\,,
\end{align}
higher order corrections in other commutators and relations appear as in the Schwarzschild case above. For example
\begin{align}
\phi_1\star\phi_1+\phi_2\star\phi_2 &\neq \phi_3\star\phi_3\,,\nn\\
\phi_4\star\phi_4+\phi_5\star\phi_5 &\neq \phi_6\star\phi_6\,,
\end{align}
which again could be interpreted as {\nc} correction to the embedding geometry.

\section{Discussion and conclusion}\label{sec:conclusion}

In this paper, we have provided explicit realizations of the Schwarzschild and the Reissner-Nordst\"om
geometry as {\nc} spaces in the framework of matrix models. Our construction is based on 
suitable embeddings of these classical geometries $\cM^4 \subset \R^D$ (``branes'') in higher-dimensional flat spaces.
These 4-dimensional branes are equipped with certain self-dual symplectic structures, which define 
the {\nc} form of these spaces via a star product. These embeddings and the corresponding
symplectic structure are chosen such that they are asymptotically constant. To be more precise, 
for $r \to \infty$ they reduce to the usual Groenewold-Moyal quantum plane which is trivially embedded 
in $\R^D$. At the semi-classical level, the central singularity is reflected by the fact that the 
embedding escapes to infinity. Non-commutative effects are expected to modify this behavior,
which is however not addressed in the present paper.

The requirement of asymptotic triviality is not satisfied by the standard embeddings
 e.g. of the 
Schwarzschild geometry in the literature \cite{Kasner:1921,Fronsdal:1959zza,Kerner:2008nt}. 
This requirement is strongly suggested by the matrix model framework, because 
the effective action may contain terms which depend on the embedding of $\cM \subset \R^D$ and
not only on its intrinsic geometry. In fact, flat 
Groenewold-Moyal quantum planes are always solutions of this class of matrix models, independent
of e.g. vacuum energy contributions. Asymptotic triviality 
is also natural since we want to consider our solution as a perturbation of 
some larger cosmological context through a localized mass. In other words, the embedding 
presented here should naturally generalize to many-particle configurations.

Another important aspect is that $e^{-\s}$, which essentially sets the scale of non-commu\-ta\-tivity,
is also asymptotically constant and non-vanishing.
This must be so because $e^{-\sigma}$ determines the strength of the non-Abelian 
gauge coupling in the matrix model \cite{Steinacker:2008ya}.
We found that $g_{\mu\nu} = G_{\mu\nu}$ (i.e. the embedding metric
coincides with the effective metric, which is certainly very natural) is indeed compatible
with asymptotically constant $\theta^{\mu\nu}$ and $e^{-\sigma}$. 
However, $e^{-\sigma}$ becomes non-trivial as one approaches 
the horizon. In fact, it turns out to vanish on a circle on the horizon, where the 
``would-be $U(1)$ gauge fields'' corresponding to $\theta^{-1}_{\mu\nu}$ vanish. 
This result, if taken literally, is somewhat problematic from the physics point of view: 
if $\theta^{\mu\nu}$ is really a rigid condensate determined by its asymptotics at 
infinity, then the rotation on the earth with respect to such a background would lead to 
small variations of the gauge coupling constant during a revolution (however
other quantities may also depend on $e^{-\sigma}$ and
lead to cancellations of such an effect). There are stringent bounds on the variations of 
the fine-structure constants \cite{rosenband}, which might exclude such an effect. 
If so, this would not rule out the framework, but it would strongly support the idea that 
the Poisson structure $\theta^{\mu\nu}$ should  be
integrated out resp. averaged over, rather than being a large-scale physical condensate.
This is indeed very natural, since  
some of the degrees of freedom in $\theta^{\mu\nu}$ 
essentially decouple from the other fields \cite{Steinacker:2010rh}. The effective action 
would then only depend on a single effective metric $G_{\mu\nu}$.
This is a very attractive possibility which will be pursued elsewhere. 

We also remind the reader that the appropriate equations governing $\theta^{\mu\nu}$ 
and therefore $e^{-\sigma}$ depend on the precise 
form of the action. $G_{\mu\nu}=g_{\mu\nu}$ is certainly natural and appropriate for
Yang-Mills models, but was simply assumed here. Relaxing this condition might also 
simplify the somewhat unusual reality properties of $\theta^{\mu\nu}$. 
There are a lot of other obvious issues arising from our construction which deserve further studies,
and the present paper should be seen as first step of a 
more general line of investigation. 
In any case, we have shown how realistic gravity can arise within this class
of matrix models, in a very explicit and accessible manner. 
This should be enough motivation for further work.

\subsection*{Acknowledgements}

H.S. would like to thank Peter Schupp and Paolo Aschieri for useful discussions.
This work was supported by the ``Fonds zur F\"orderung der Wissenschaftlichen Forschung'' 
(FWF) under contract P21610-N16.

\startappendix
\Appendix{Derivation of the symplectic form \eqnref{eq:symplectic-solution}}\label{sec:appendix-symplectic}
Considering \eqnref{eq:syf-ansatz} we can make the additional ansatz that $\syf$ is invariant under the Killing vector fields, i.e. $\cL_{V_{ts}}\syf = 0$, and moreover $\cL_{V_{ts}} \star = \star \cL_{V_{ts}}$.
This implies $dE = 0 = \cL_{V_{ts}}E$, and together with  $\cL_{V_\vph} E = 0$ we obtain
\be
E =  E_r(r,\vth) dr + E_\vth(r,\vth) d\vth
 = d \chi_E(r,\vth)\,. 
\ee
Similarly, $\cL_{V_\varphi}\syf = 0$ implies $\syf_B = B \wedge d\varphi$ with 
\be
B = B_r d\vth+B_\vth dr = d\chi_B(r,\vth)\,.
\ee
Now we need to work out 
\be
\star \syf
= \frac 1{2\sqrt{|g|}}g_{\a\a'}g_{\b\b'}\varepsilon^{\a'\b'\mu\nu}
\theta^{-1}_{\mu\nu} dx^\a\wedge dx^\b\,,
\ee
where in Schwarzschild coordinates
\begin{align}
g_{tt} &= (1-\frac{r_c}{r}), 
 && g_{rr} = (1-\frac{r_c}{r})^{-1}\,, \nn\\
g_{\vth\vth} &= r^2, 
&& g_{\varphi\varphi} = r^2 \sin^2\vth, \nn\\
\sqrt{|g|} &= r^2 \sin\vth\,.
\end{align}
So if we define
\begin{align}
\syf_B &:= \star \syf_E 
= \star(E_r dr + E_\vth d\vth)\wedge dt\nn\\
&= \frac 1{r^2\sin\vth}\,
\left(r^4 \sin^2\vth E_r d\vth d\varphi 
-  r^2(1-\frac{r_c}{r})^{-1}\sin^2\vth E_\vth d r d\varphi\right)\nn\\
&= \sin\vth\,
\left(r^2 E_r d\vth 
-  (1-\frac{r_c}{r})^{-1} E_\vth d r \right)
\wedge d\varphi  \nn\\
&= (B_r d\vth + B_\vth dr)\wedge d\varphi\,,
\end{align}
and if that is closed, then $\syf = i \syf_E + \star \syf_E $ is self-dual. 
Explicitly, for
\be
E = d(f(r) \cos\vth) = f' \cos\vth dr -f\sin\vth d\vth
= E_r dr + E_\vth d\vth
\ee
we need
\begin{align}
0 &= d\left(r^2 E_r \sin\vth\,d\vth 
-  (1-\frac{r_c}{r})^{-1} E_\vth \sin\vth\,d r \right) \nn\\
 &= d\left(r^2 f'\sin\vth\,\cos\vth d\vth 
+ f(1-\frac{r_c}{r})^{-1} \sin^2\vth d r \right) \nn\\
 &= \del_r(r^2 f')\sin\vth\,\cos\vth dr \wedge d\vth 
+ 2f(1-\frac{r_c}{r})^{-1} \sin\vth\cos\vth d\vth\wedge d r\,,
\end{align}
so
\begin{align}
\del_r(r^2 f') &=  2f(1-\frac{r_c}{r})^{-1} \,, \nn\\
r^2 f'' +  2 r f' - 2 f (1-\frac{r_c}{r})^{-1} &= 0\,,
\end{align}
which has the solution
\be
f(r) = c_1 r(1-\frac{r_c}{r})  + c_2 \frac 1{r_c^2}
    \(1-\frac{r_c}{2r} + (\frac{r}{r_c}-1) \ln(1-\frac{r_c}r)\)\,.
\ee
For $c_2=0$ we get \eqnref{eq:symplectic-solution} 
which has the desired asymptotics as an
asymptotically constant external  field. 
Then 
\begin{align}
\sqrt{|\theta^{-1}|} &= {\rm Pfaff}(\theta^{-1}_{\mu\nu})
= \inv{8} \varepsilon^{\mu\nu\rho\sigma}\theta^{-1}_{\mu\nu} \theta^{-1}_{\rho\sigma} \nn\\
&= ( \theta^{-1}_{r t}\theta^{-1}_{\vth\varphi} - \theta^{-1}_{\vth t}\theta^{-1}_{r\varphi} ) \nn\\
&= (E_r B_r - E_\vth B_\vth) \nn\\
&= \left(f'^2 r^2\cos^2\vth \sin\vth + f^2\sin^3\vth(1-\frac{r_c}{r})^{-1}\right)\nn\\
&= \cst^2r^2\sin\vth\left(1-\sin^2\vth\frac{r_c}{r}\right)\,,
\label{theta-det}
\end{align}
which yields \eqref{eq:esig-solution}.

\Appendix{Commutation relations for Schwarzschild geometry}\label{sec:appendix-7dim-poisson}
From \eqnref{eq:darboux} we can immediately read off $\th^{-1}_{\m\n}$ in Darboux coordinates, and its inverse leads to the Poisson brackets
\begin{align}
\pb{x_D^\m}{x_D^\n}&=\sth\left(\begin{array}{cccc}
0 & i & 0 & 0\\
-i & 0 & 0 & 0\\
0 & 0 & 0 & 1\\
0 & 0 & -1 & 0
\end{array}\right)\,,
\end{align}
where $\sth=1/\cst$. Using the relations
\begin{align}
H_{ts}&=r\g\cos\vth=z\left(1-\frac{r_c}{r}\right)\,,\nonumber\\
H_\vph&=\inv{2}r^2\sin^2\vth=\inv{2}\left(x^2+y^2\right)\,,\nonumber\\
t_S&=t-r_c\ln\left|\frac{r}{r_c}-1\right|\,,\nonumber\\
r&=\sqrt{x^2+y^2+z^2}\,,
\end{align}
we transform the set of coordinates to $\{H_{ts},t_S,H_\vph,\vph\}\to\{t,x,y,z\}$, and get
\begin{align}
\pa_{H_{ts}}&=e^\bs\left(\frac{r_cz}{r\left(r-r_c\right)}\pa_t+\pa_z\right)
\,,\nn\\
\pa_{t_S}&=\pa_t\,,\nn\\
\pa_{H_\vph}&=e^\bs\left(\frac{r_c}{r^2}\pa_t-\frac{r_cz}{r^3}\pa_z\right)+\inv{x^2+y^2}\left(x\pa_x+y\pa_y\right)
\,,\nn\\
\pa_\vph&=-y\pa_x+x\pa_y\,,
\label{eq:app-coordinate-trafo}
\end{align}
where $e^{-\bs}=\sth e^{-\s}$ is the Jacobian determinant of the transformation. 
We hence arrive at the following Poisson brackets in terms of the Cartesian coordinates $x^\m=\{t,x,y,z\}$:
\begin{align}
\pb{x^\m}{x^\n}&=\sth e^\bs \left(\begin{array}{cccc}
0 & -\frac{r_cy}{r^2} & \frac{r_cx}{r^2} & -i  \\
 \frac{r_cy}{r^2} & 0 & e^{-\bs} & -\frac{r_cyz}{r^3}  \\
 -\frac{r_cx}{r^2} & -e^{-\bs} & 0 & \frac{r_cxz}{r^3} \\
 i & \frac{r_cyz}{r^3} & -\frac{r_cxz}{r^3} & 0 
\end{array}\right)\,.
\end{align}
Using these, one easily works out the remaining Poisson brackets with the embedding functions $\phi_i$ of \eqnref{eq:embedding-fcts}, namely $\pb{x^\m}{\phi_i(x)}$ and $\pb{\phi_i(x)}{\phi_j(x)}$, leading finally to \eqref{eq:theta7-dir_c}.

\paragraph{Next-to-leading order commutation relations:}
To third order in the expansion parameter $\sth$ one finds the star commutators
\begin{align}
\starco{t}{x}&=- i\sth\frac{r_c e^\bs}{r^2}y-\sth^3yF_{txy}+\mO{\sth^5}\,,\nonumber\\
\starco{t}{y}&= i\sth\frac{r_c e^\bs}{r^2}x+\sth^3xF_{txy}+\mO{\sth^5}\,,\nonumber\\
\starco{t}{z}&=\sth e^\bs\,,\nonumber\\
\starco{x}{y}&= i\sth\,,\nonumber\\
\starco{x}{z}&=- i\sth y\frac{r_cz e^\bs}{r^3}-\sth^3yF_{zxy}+\mO{\sth^5}
\,,\nonumber\\
\starco{y}{z}&=- i\sth x\frac{r_cz e^\bs}{r^3}+\sth^3xF_{zxy}+\mO{\sth^5}
\,,
\label{eq:nlo-co-rel-txyz}
\end{align}
with the abbreviations
\begin{align}
F_{txy}&= \frac{r_c e^{5\bs}}{24 r^{14}}\bigg( \g ^2 r^6 (3 r_c^2 - 9 r_c r + 8 r^2) - \g  r_c r^4 (6 \g  r_c + 17 r) z^2 \nonumber\\*
&\qquad\qquad + r_c^2 (3 r_c^2 + 3 r_c r + 2 r^2) z^4\bigg) \,, \nonumber\\
F_{zxy}&= \frac{r_cz e^{5\bs}}{8 r^{15}} \bigg(\g ^2 r^6 (r_c^2 - 4 r_c r + 5 r^2) + 2 \g  (\g ^2-3) r_c r^5 z^2 + r_c^4 z^4\bigg) \,.
\end{align}
Notice, that some expressions (i.e. the ones where $\mO{\sth^n}$ is omitted) are exact to all orders\footnote{However, while this is the case for the commutator $\starco{x}{y}= i\sth$, the according anticommutator does in fact have higher order contributions, i.e. \[\staraco{x}{y}=2xy-xy\left(\frac{\sth^2}{(x^2+y^2)^2}-\frac{\sth^4}{4(x^2+y^2)^4}\right)+\mO{\sth^6}\,.\]}. \\
Furthermore, we find for the embedding functions $\phi_i$ to third order in $\sth$:
{\allowdisplaybreaks
\begin{align}
\starco{t}{\phi_1}&= -\sth e^\bs\frac{z}{r}f^+_{12}(0)+\mO{\sth^5}=-\sth e^\bs\frac{z}{r}\left(\frac{\phi_1}{2r}+\w\phi_2\right)+\sth^3\phi_2 F_{t\phi12}+\mO{\sth^5}\,,\nonumber\\
\starco{t}{\phi_2}&= -\sth e^\bs\frac{z}{r}f^-_{21}(0)+\mO{\sth^5}=-\sth e^\bs\frac{z}{r}\left(\frac{\phi_2}{2r}-\w\phi_1\right)-\sth^3\phi_1 F_{t\phi12}+\mO{\sth^5}\,,\nonumber\\
\starco{t}{\phi_3}&= -\sth e^\bs\frac{z\phi_3}{2r^2}\,,\nonumber\\
\starco{x}{\phi_1}&= - i\sth e^\bs\frac{y}{r}f^+_{12}(r_c)+\sth^3yF_{xy}(\phi_1,\phi_2)+\mO{\sth^5}\,,\nonumber\\
\starco{x}{\phi_2}&= - i\sth e^\bs\frac{y}{r}f^-_{21}(r_c)+\sth^3yF_{xy}(-\phi_2,\phi_1)+\mO{\sth^5}\,,\nonumber\\
\starco{x}{\phi_3}&= - i\sth e^\bs\frac{y\g \phi_3}{2r^2}+\sth^3yF_{\phi3xy}+\mO{\sth^5}\,,\nonumber\\
\starco{y}{\phi_1}&=  i\sth e^\bs\frac{x}{r}f^+_{12}(r_c)-\sth^3xF_{xy}(\phi_1,\phi_2)+\mO{\sth^5}\,,\nonumber\\
\starco{y}{\phi_2}&=  i\sth e^\bs\frac{x}{r}f^-_{21}(r_c)-\sth^3xF_{xy}(-\phi_2,\phi_1)+\mO{\sth^5}\,,\nonumber\\
\starco{y}{\phi_3}&=  i\sth e^\bs\frac{x\g \phi_3}{2r^2}-\sth^3xF_{\phi3xy}+\mO{\sth^5}\,,\nonumber\\
\starco{z}{\phi_1}&= \sth e^\bs\w\phi_2+\sth^3\phi_2 F_{z\phi12}+\mO{\sth^5}\,,\nonumber\\
\starco{z}{\phi_2}&= -\sth e^\bs\w\phi_1-\sth^3\phi_1 F_{z\phi12}+\mO{\sth^5}\,,\nonumber\\
\starco{z}{\phi_3}&= 0\,,
\end{align}
with
\begin{align}
F_{t\phi12}&= \frac{r_cz\w^3 e^{2\bs}}{24 \g ^3 r^6} \Big(27  e^{3\bs} (z^2+r^2\g (1+\g )) -6 \g  r^2 + 
     e^{\bs} (3 (7 + \g  (9 + 5 \g )) r^2 + 2 z^2) \nonumber\\
&\quad\qquad - 3  e^{2\bs} \left((9 + 2 \g  (8 + 7 \g )) r^2 + 7 z^2\right)\Big) \,,\nonumber\\
F_{\phi3xy}&=  i e^{5\bs}\frac{\g  \phi_3}{64 r^{10}}\Big(15 \g ^4 r^4 - 2 \g  \left(8 + \g  \left(9 + \g  (15 \g -32)\right)\right) r^2 z^2 \nonumber\\
&\quad\qquad + (1 - \g )^2 \left(20 + \g  (15 \g -34)\right) z^4\Big)\,,\nonumber\\
F_{z\phi12}
&=\frac{\w^3 e^{3\bs}}{8 r^4} ( e^\bs-1) \Big((9  e^\bs-4) z^2 - r^2\Big) \,,
\end{align}
and
\begin{small}
\begin{align}
&F_{xy}(\phi_i,\phi_j)=-\frac{ i  e^{5\bs} }{192 r^{14}}\Big(8  e^{-2 \bs } \w^3 \phi_j r^{11}-12 \w^2 \phi_i r^3 \left[(r_c-3 r) \g ^2 r^6+2 r_c^2 z^2  \g  r^3+r_c^2 (r_c+3 r) z^4\right]\nonumber\\*
& -2 \w \phi_j r^4 \left[\left(15 r_c^2-40 r r_c+33 r^2\right) \g ^2 r^3+2 r_c \left(15  r_c^2-16 r r_c-33 r^2\right) z^2 \g+r_c^2 \left(15 r_c^2+8 r r_c+9 r^2\right) \tfrac{z^4}{r^3}\right]\nonumber\\*
& -3 \g  \phi_i \left[15 \g ^4 r^8+2 r_c \left(15 r_c^2-13 r r_c-10 r^2\right) z^2 \g  r^3+r_c^2 \left(15 r_c^2+4 r r_c+r^2\right) z^4\right]\Big)
\,.
\end{align}
\end{small}
Finally we also have
\begin{align}
\starco{\phi_1}{\phi_2}&= \sth e^\bs\frac{\w z\phi_3^2}{2r^2}
+\sth^3 e^{3\bs}\frac{\w^3 z \phi_3^2}{16 r^6}\Big(4  e^\bs (r^2 + z^2)-r^2 - 9 z^2 e^{2\bs}\Big)
+\mO{\sth^5},\nonumber\\
\starco{\phi_1}{\phi_3}&= \sth e^\bs\frac{\w z\phi_3\phi_2}{2r^2}+\sth^3\phi_2F_{\phi312}+\mO{\sth^5}\,,\nonumber\\
\starco{\phi_2}{\phi_3}&= -\sth e^\bs\frac{\w z\phi_3\phi_1}{2r^2}-\sth^3\phi_1F_{\phi312}+\mO{\sth^5}\,,
\label{eq:nlo-co-rel-phi}
\end{align}
with
\begin{align}
F_{\phi312}&= e^{3\bs}\frac{\w^3 z \phi_3}{64 r^6}\Big((1 - 22  e^\bs + 36  e^{2\bs}) (x^2 + y^2)-(7 - 38  e^\bs + 36  e^{2\bs}) r^2\Big)\,.
\end{align}
}



\end{document}